\documentstyle[10pt,epsf,epsfig,dp_delphititle,graphicx]{dp_delphi}
\newcommand{\mha}       {\mbox{($m_{\mathrm{h}}$,$m_{\mathrm{A}}$)}}
\newcommand{\mhtgb}       {\mbox{($m_{\mathrm{h}}$,$ \tan\beta $)}}
\newcommand{\matgb}       {\mbox{($m_{\mathrm{A}}$,$ \tan\beta $)}}
\newcommand{\mtop} {\mbox{$m_{\mathrm{t}}$}}
\newcommand{\ra}    {\rightarrow}
\newcommand{\tanb}  {\mbox{$ \tan\beta $}}

\newcommand{\mh}        {\mbox{$m_{\mathrm{h}}$}}

\makeindex
\begin{document}


%
\pagenumbering{arabic}                              
\renewcommand{\thefootnote}{\fnsymbol{footnote}}    
\setcounter{footnote}{1}                            %
\large

\section*{A More General MSSM Parameter Scan}

Here a more general MSSM parameter scan than that described in section 8.3
is considered, following~\cite{jras,as97}.
In addition, while the benchmark results given there used 
the Renormalisation Group Equation (RGE) approach~\cite{RGE} 
with recent modifications bringing them into agreement 
with the Full Diagrammatic Calculations (FDC)~\cite{holliknew}, 
here the FDC are used directly.

The parameters shown in 
Table~\ref{tab:allparameters} are the input parameters 
for the calculations of the physical Higgs, sfermion, chargino, and
neutralino masses and production rates, and they are varied in the ranges shown.
The A mass is scanned in steps of 1~${\rm GeV/c^2}$
up to 200~${\rm GeV/c^2}$ and in larger steps thereafter.
For each $m_{\mathrm{A}}$, 2700 parameter combinations are investigated.
The range of \tanb\ is restricted to $0.5\leq\tanb\leq 50$:
the lower bound is described for example in~\cite{GRGU},
and affects the theoretically allowed regions in the \mha\ plane;
varying the upper bound has no significant effect on the results.
The range of $\mu$ is restricted to $\rm \pm 500~GeV/c^2$,
since very large values are disfavored by Charge- and Color-Breaking (CCB)
global minima in the MSSM~\cite{ccb}.
Finally, in the RGE approach used for the benchmark scan, 
$X_t = \sqrt{6} M_S$ defined the so-called $m_{\mathrm h}^{max}$ scenario, 
and $X_t = 0$ the no mixing scenario. 
In the FDC used here, the maximum scalar Higgs mass is obtained for $A=2M_S$,
where $X^{FDC}_t = A - \mu / \tan\beta$.
Compared to previous similar studies~\cite{jras,as97}, 
the range of $A$ is extended from $\pm 1M_S$ to $\pm 2M_S$ 
in order to include the maximal Higgs boson mass in the scan.
The top quark mass is fixed at $\mtop= 175 \, {\rm GeV/c^2}$.
A value of $\mtop= 180 \, {\rm GeV/c^2}$ would increase the calculated
$m_{\mathrm h}$ values by 2~${\rm GeV/c^2}$ for large $\tan\beta$ 
and by 4~${\rm GeV/c^2}$ for small $\tan\beta$.

\begin{table}[htb]
\vspace*{-0.4cm}
\begin{center}
\begin{tabular}{|c|c|c|c|c|c|c|}
\hline
Parameter & $m_{\mathrm{A}}$ & \tanb & $M_S$ 
& $M_{\mathrm{2}}$  & $\mu$ &$A$ or $X_t$\\
\hline
this scan     & 20\,---\,1000 & $0.5$\,---\,50 & 200\,---\,1000 &
200\,---\,1000
& $-500$\,---\,+500 & $-2M_S$\,---\,$+2M_S$ \\ 
benchmark  & 20\,---\,1000 & $0.5$\,---\,50 &1000&200&-200&0,~$\sqrt{6}M_S$\\
\hline
\end{tabular}
\end{center}
\vspace*{-0.4cm}
\caption[]{\label{tab:allparameters}
\baselineskip=12pt
Ranges of SUSY parameters at the electroweak scale used for independent
variation in the more general MSSM parameter scan compared with those used 
in the benchmark scan of section 8.3. All masses are given in ${\rm GeV/c^2}$.}
\end{table}

For some parameter combinations the branching ratio into a pair
of neutralinos is dominant. In such a case no limit can be derived
using the channels studied in this paper. Therefore, limits from the
DELPHI search for invisible
decays of neutral Higgs bosons~\cite{inv} are also used.

The following constraints have also been considered:
\begin{itemize}
\item The rate of the flavour changing neutral current process
${\mathrm b} \ra {\mathrm s}\gamma$~\cite{cleo}. 
In the SM only amplitudes with virtual ${\rm W}^\pm$ 
exchange contribute, while in the MSSM there 
are additional contributions from supersymmetric particles and Higgs 
bosons~\cite{ciuchini1}\footnote{These calculations changed the expected
b $\ra$ s$\gamma$ rates significantly. However, not all contributing 
terms are included and resulting assumptions are not valid in this
more general scan.}.
\item The electroweak parameter
$\Delta \rho = \alpha_{\rm em}T_{\rm MSSM}$~\cite{chank}. The MSSM contributions
are constrained by experiment: 
$T_{\rm MSSM} < 0.08$~\cite{erler}.
\item Chargino and neutralino mass limits from direct searches~\cite{SUSY}.
\end{itemize}
But these constraints have little influence on the excluded
parameter regions with the present Higgs mass limits.
Therefore, the excluded parameter sets are determined from the Higgs boson 
searches alone.

\begin{figure}[H]
  \begin{center}
    \includegraphics[width=\textwidth]{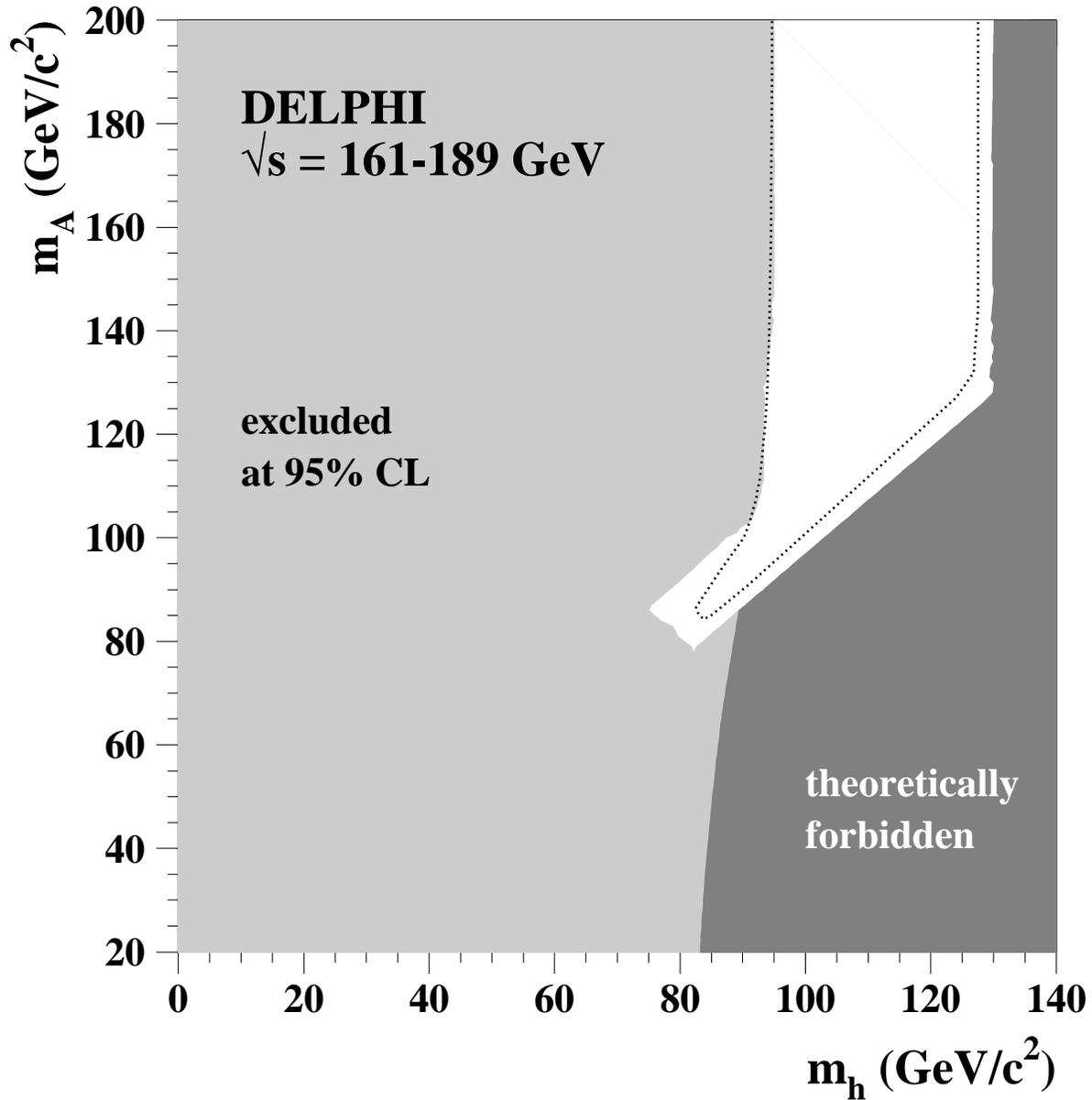}
  \end{center}
  \caption[]{\label{fig:mh_ma}
           MSSM exclusion based on all DELPHI data from 
           $\sqrt{s} = 161$~GeV to 189~GeV
           in the framework of the more general parameter scan.
           The region excluded at 95\% CL (light grey),
           the unexcluded region (white) and the theoretically forbidden
           region (dark grey) are shown. The dotted line marks the allowed
           region obtained with the
           benchmark scan under the assumption of maximal mixing in
           the stop sector.
           }
\end{figure}

\begin{figure}[H]
  \begin{center}
    \includegraphics[width=\textwidth]{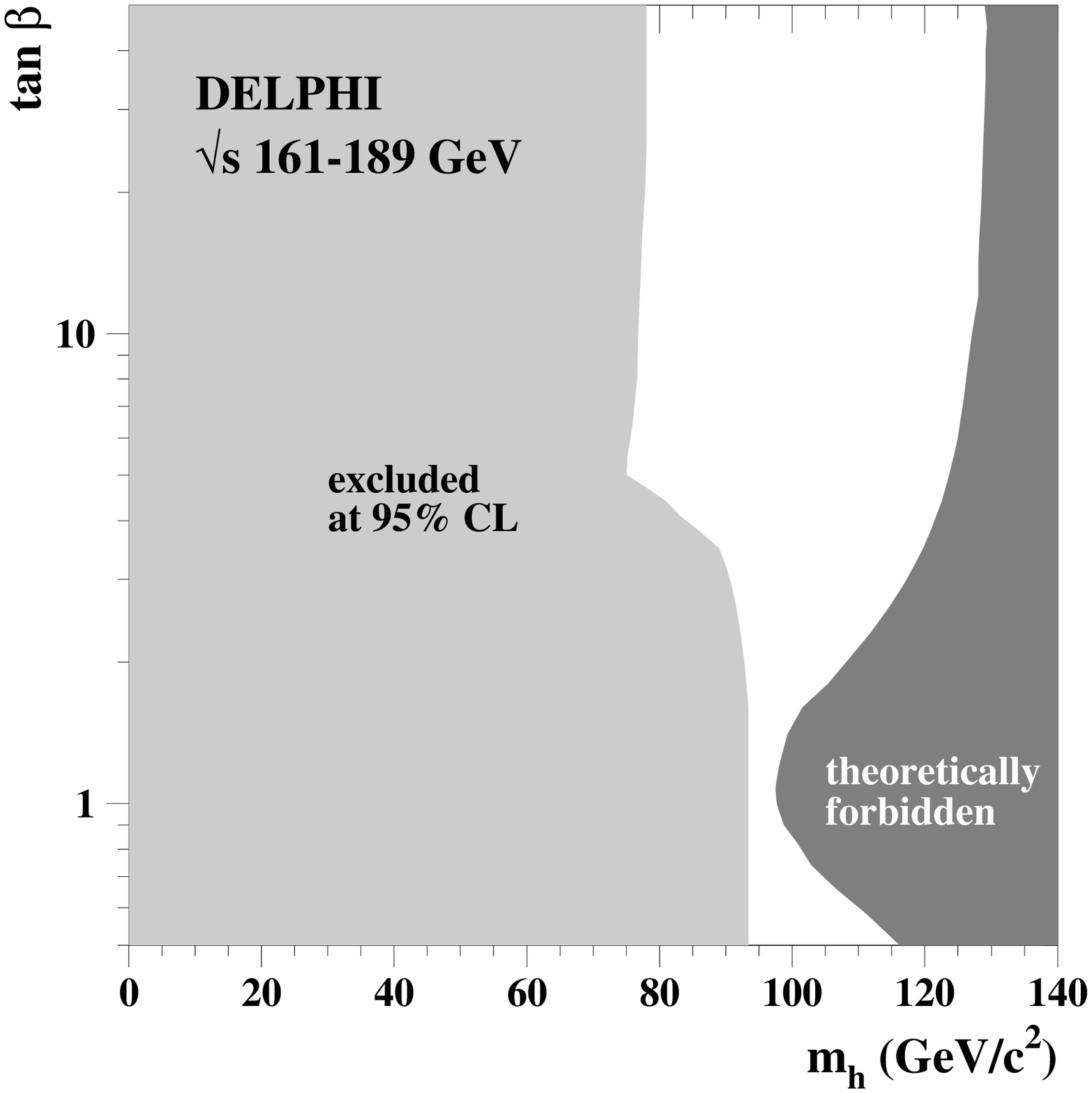}
  \end{center}
  \caption[]{\label{fig:mh_tgb}
           MSSM exclusion based on all DELPHI data from 
           $\sqrt{s} = 161$~GeV to 189~GeV
           in the framework of the more general parameter scan.
           The region excluded at 95\% CL (light grey),
           the unexcluded region (white) and the theoretically forbidden
           region (dark grey) are shown.
           }
\end{figure}

\begin{figure}[H]
  \begin{center}
    \includegraphics[width=\textwidth]{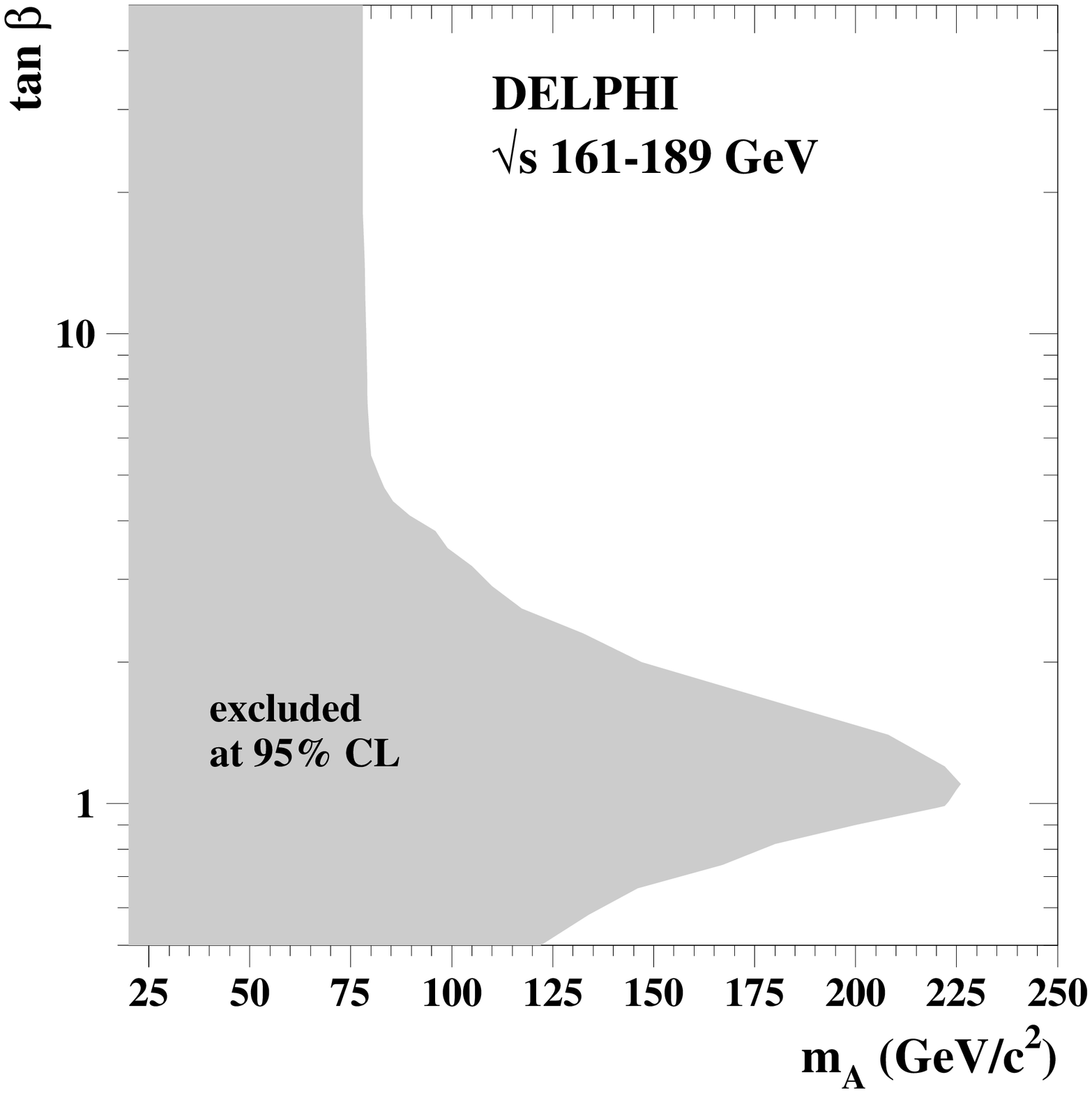}
  \end{center}
  \caption[]{\label{fig:ma_tgb}
           MSSM exclusion based on all DELPHI data from 
           $\sqrt{s} = 161$~GeV to 189~GeV
           in the framework of the more general parameter scan.
           The region excluded at 95\% CL (light grey) and
           the unexcluded region (white) are shown.
           }
\end{figure}

Figures~\ref{fig:mh_ma} to~\ref{fig:ma_tgb} present the results 
in the \mha, \mhtgb, and \matgb\ planes respectively.
A given \mha\ combination, for example, is excluded if all
contrubuting SUSY parameter sets in the ranges defined in
Table~\ref{tab:allparameters} are excluded 
at more than 95\% confidence level 
after combining all search channels
using the likelihood ratio method~\cite{alex}.
The figures show three regions:
\begin{itemize}
\item the 95\% CL excluded region (light grey),
\item the theoretically not allowed region (dark).
\item the allowed region (white),
\end{itemize}
and Fig.~\ref{fig:mh_ma} also shows the region excluded by the benchmark scan 
(dotted line):
the region excluded by the more general scan is smaller.
In particular, the benchmark limits of 82.6~${\rm GeV/c^2}$ on the h mass and 
84.1~${\rm GeV/c^2}$ on the A mass are reduced to 75~${\rm GeV/c^2}$ 
and 78~${\rm GeV/c^2}$, respectively.
Comparison of the FDC cross sections used here and the RGE calculations 
used for the benchmark scan confirms that 
this is due to the reduced production cross-sections
allowed by the extended parameter range, not due to differences between the
FDC and RGE calculations.
As illustrated in Table~\ref{tab:unexcl},
low unexcluded \mh\ values are typically obtained for large mixing 
in the stop sector (large $A$, large $|\mu|$).
In conclusion, the scan over a larger parameter region reduces the mass limits 
given in sections 8.3 and 9 
by 8~${\rm GeV/c^2}$ for the scalar and
by 6~${\rm GeV/c^2}$ for the pseudoscalar
Higgs boson.

\begin{table}[hp]
\vspace*{-0.4cm}
\begin{center}
\begin{tabular} {|c|c|c|c|c|c|c|c|c|c|c|c|}\hline
$m_{\rm A}$ & $m_{\rm h}$ & $\tan\beta$ & $M_S$ & $M_{\rm 2}$ &
$\mu$ & $A/M_S$ & $X^{FDC}_t$ &
$m_{\tilde{\rm t}1}$ & $m_{\tilde{\rm t}2}$ & $\sigma_{\rm hZ}^{189}$ &
$\sigma_{\rm hA}^{189}$ \\
\hline
80  & 78  &   8 & 1000 & 200  &  500 &  2 &  1938 & 979 & 1048 & 0.07 &
0.09 \\
85  & 80  &  12 & 200  & 1000 & -500 &  1 &   242 & 162 &  333 & 0.00 &
0.06 \\
85  & 86  &  10 & 1000 & 1000 &  500 &  2 &  1950 & 824 & 1174 & 0.03 &
0.04 \\
90  & 86  & 4.4 & 1000 & 200  & -100 & -2 & -1977 & 824 & 1174 & 0.11 &
0.02 \\
\hline
\end{tabular}
\end{center}
\vspace*{-0.4cm}
\caption[]{\label{tab:unexcl}
Examples of unexcluded parameter combinations in the more general MSSM scan.
Cross-sections for Higgs boson bremsstrahlung and pair-production
are given for $\sqrt{s} = 189$~GeV.
All masses are given in GeV/$c^2$ and cross-sections in pb.}
\end{table}


\newcommand{\group}[2]{#1 Collaboration, #2 \etal }
\newcommand{\us   }   {\group{L3}{B.~Adeva}}
\newcommand{\usnew}   {\group{L3}{O.~Adriani}}
\newcommand{\etal }   {{\em et~al.}}


\clearpage

\clearpage

\clearpage

\end{document}